\documentclass[aps,prl,twocolumn,showpacs,amsmath,amssymb,superscriptaddress]{revtex4}
\pdfoutput=1
\usepackage[utf8]{inputenc}
\usepackage{amsmath}
\usepackage{bm}
\usepackage{booktabs}
\usepackage{siunitx}
\usepackage{graphicx}
\usepackage{multirow}
\usepackage{hyperref}
\usepackage{float}
\usepackage[version=4]{mhchem}
\begin{document}

\title{Neutron Spin Resonance Near a Lifshitz Transition in Overdoped Ba$_{0.4}$K$_{0.6}$Fe$_2$As$_2$}

\author{Yang Li}
\affiliation{Beijing National Laboratory for Condensed Matter Physics, Institute of Physics, Chinese Academy of Sciences, Beijing 100190, China}
\affiliation{School of Physical Sciences, University of Chinese Academy of Sciences, Beijing 100190, China}
\author{Dingsong Wu}
\affiliation{Beijing National Laboratory for Condensed Matter Physics, Institute of Physics, Chinese Academy of Sciences, Beijing 100190, China}
\affiliation{School of Physical Sciences, University of Chinese Academy of Sciences, Beijing 100190, China}
\affiliation{Department of Physics, University of Oxford, Oxford OX1 3PU, United Kingdom}
\author{Yingjie Shu}
\affiliation{Beijing National Laboratory for Condensed Matter Physics, Institute of Physics, Chinese Academy of Sciences, Beijing 100190, China}
\affiliation{School of Physical Sciences, University of Chinese Academy of Sciences, Beijing 100190, China}
\author{Bo Liu}
\affiliation{Beijing National Laboratory for Condensed Matter Physics, Institute of Physics, Chinese Academy of Sciences, Beijing 100190, China}
\affiliation{School of Physical Sciences, University of Chinese Academy of Sciences, Beijing 100190, China}
\author{Uwe Stuhr}
\affiliation{Laboratory for Neutron Scattering and Imaging, Paul Scherrer Institut, CH-5232 Villigen PSI, Switzerland}
\author{Guochu Deng}
\affiliation{Australian Centre for Neutron Scattering, Australian Nuclear Science and Technology Organisation, Lucas Heights NSW-2234, Australia}
\author{Anton P. J. Stampfl}
\affiliation{Australian Centre for Neutron Scattering, Australian Nuclear Science and Technology Organisation, Lucas Heights NSW-2234, Australia}
\author{Lin Zhao}
\affiliation{Beijing National Laboratory for Condensed Matter Physics, Institute of Physics, Chinese Academy of Sciences, Beijing 100190, China}
\affiliation{School of Physical Sciences, University of Chinese Academy of Sciences, Beijing 100190, China}
\author{Xingjiang Zhou}
\affiliation{Beijing National Laboratory for Condensed Matter Physics, Institute of Physics, Chinese Academy of Sciences, Beijing 100190, China}
\affiliation{School of Physical Sciences, University of Chinese Academy of Sciences, Beijing 100190, China}
\author{Shiliang Li}
\affiliation{Beijing National Laboratory for Condensed Matter Physics, Institute of Physics, Chinese Academy of Sciences, Beijing 100190, China}
\affiliation{School of Physical Sciences, University of Chinese Academy of Sciences, Beijing 100190, China}
\author{Amit Pokhriyal}
\affiliation{Theory and Computational Physics Section, Raja Ramanna Centre for Advanced Technology, Indore 452013, India}
\affiliation{Homi Bhabha National Institute, BARC training school complex 2nd floor, Anushakti Nagar, Mumbai 400094, India}
\author{Haranath Ghosh}
\email{hng@rrcat.gov.in}
\affiliation{Theory and Computational Physics Section, Raja Ramanna Centre for Advanced Technology, Indore 452013, India}
\affiliation{Homi Bhabha National Institute, BARC training school complex 2nd floor, Anushakti Nagar, Mumbai 400094, India}
\author{Wenshan Hong}
\email{wshong@iphy.ac.cn}
\affiliation{Beijing National Laboratory for Condensed Matter Physics, Institute of Physics, Chinese Academy of Sciences, Beijing 100190, China}
\author{Huiqian Luo}
\email{hqluo@iphy.ac.cn}
\affiliation{Beijing National Laboratory for Condensed Matter Physics, Institute of Physics, Chinese Academy of Sciences, Beijing 100190, China}

\date{\today}

\begin{abstract}

Elucidating the relationship between spin excitations and fermiology is essential for clarifying the pairing mechanism in iron-based superconductors (FeSCs). Here, we report inelastic neutron scattering results on the hole overdoped Ba$_{0.4}$K$_{0.6}$Fe$_2$As$_2$ near a Lifshitz transition, where the electron pocket at $M$ point is nearly replace by four hole pockets. In the normal state, the spin excitations are observed at incommensurate wave vectors with chimney-like dispersions. By cooling down to the superconducting state, a neutron spin resonance mode emerges with a peak energy of $E_r=$ 14-15 meV weakly modulated along $L$-direction. The incommensurability notably increases at low energies, giving rise to downward dispersions of the resonance mode. This behavior contrasts sharply with the upward dispersions of resonance observed in optimally doped Ba$_{0.67}$K$_{0.33}$Fe$_2$As$_2$ contributed by the hole to electron scattering, but resembles with the cases in KFe$_2$As$_2$ and KCa$_2$Fe$_4$As$_4$F$_2$ where the fermiology are dominated by hole pockets. These results highlight the critical role of electronic structure modifications near the Fermi level, especially in governing interband scattering under imperfect nesting conditions, which fundamentally shape the spin dynamics of FeSCs.

\end{abstract}

\maketitle

Understanding the superconductivity and related phenomena in iron-based superconductors (FeSCs) is a great challenge due to their complex multiband nature\cite{Scalapino2012}. Such complexities arise from multiple Fermi surfaces, diverse superconducting gap structures, intricate orbital contributions, and prominent spin fluctuations, particularly those stemming from interband scattering\cite{Fernandes2022}. These factors become especially significant when comparing iron-based superconductors with different Fermi surface topologies. In most FeSCs, the Fermi surface comprises multiple hole pockets at the $\Gamma$ point and hybridized electron pockets at the $M$ point, facilitating superconductivity with $s_{\pm}$ pairing symmetry driven by spin fluctuations. These fluctuations may originate either from weak-coupling Fermi surface nesting contributed by itinerant electrons, or from strong-coupling local magnetic interactions on the Fe sites\cite{Mazin2008, Kuroki2008, Si2008, Seo2008, Dai2015}. In contrast, systems with only electron-like Fermi pockets typically observed in electron-doped iron chalcogenides exhibit dominant $d$-wave pairing symmetry \cite{Zhang2011, Qian2011, Liu2012, Zhao2016, Wo2025}, driven by inter-pocket scattering between zone corners\cite{Maier2011,Wang2011}. The spin excitations in such systems are distinct between the parent and superconducting phases\cite{Dai2012, Wang2013, Pan2017, Friemel2012, Wang2012, Wo2025}, suggesting that the specific multiband Fermi surface topologies strongly influence the nature of superconductivity and quasi-particle excitations.

\begin{figure}[htbp] \centering
\includegraphics[width=0.45\textwidth]{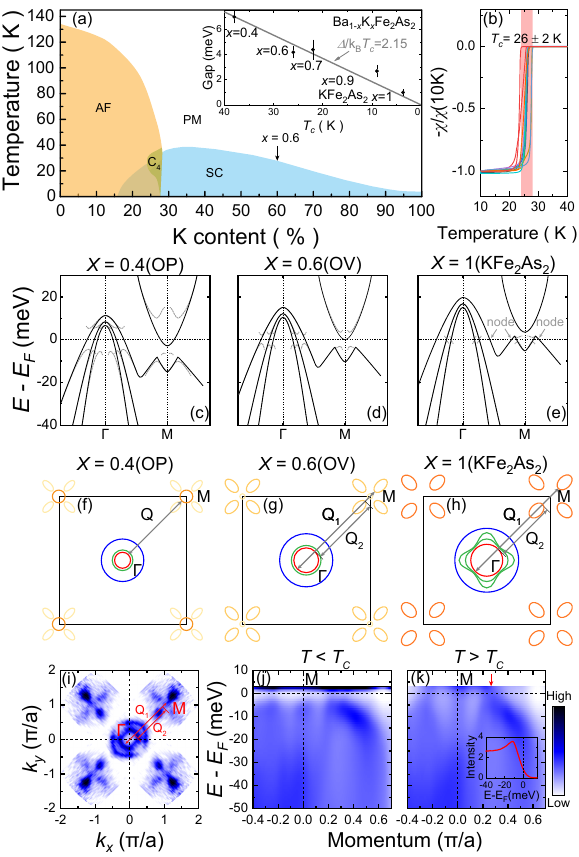}
\caption{Phase diagram and electronic structure of Ba$_{1-x}$K$_x$Fe$_2$As$_2$. (a) Phase diagram of Ba$_{1-x}$K$_x$Fe$_2$As$_2$\cite{Bohmer2015}, with the arrow indicating the doping level of $x=0.6$. The inset shows the superconducting gap size near $\Gamma$ point\cite{Wu2024,Xu2013,Nakayama2011,Cai2021}. (b) Zero-field-cooled magnetization measurements from multiple crystals, yielding an average $T_c$ of 26 K. (c)-(e) Schematic representations of the band structure for Ba$_{1-x}$K$_x$Fe$_2$As$_2$ at $x = 0.4$ (optimally doped), $x = 0.6$ (overdoped, this study), and $x = 1$ (KFe$_2$As$_2$, full overdoped). The gray dashed line marks the Bogoliubov back-bending effect due to the superconducting gap. (f)-(h) Corresponding Fermi surface illustrations, highlighting interband scattering vectors (${\boldsymbol{Q}}$, ${\boldsymbol{Q_1}}$ and ${\boldsymbol{Q_2}}$). (i)-(k) ARPES measurements of the Fermi surfaces and band structure for $x=0.6$. The inset in (k) shows the energy distribution curve at the position marked by a arrow.}
\label{figure1}
\end{figure}

Ba$_{1-x}$K$_x$Fe$_2$As$_2$ is a typical hole-doped FeSC with a broad doping range, from the parent BaFe$_2$As$_2$ to the fully hole doped KFe$_2$As$_2$ \cite{Bohmer2015,Wu2024,Xu2013,Nakayama2011,Cai2021,Kihou2016} (Fig.~\ref{figure1}(a) and (b)). Intriguing phenomena in the overdoped region are argued to be related to a Lifshitz transition of Fermi surfaces \cite{Thomale2011,Grinenko2020,Bartl2025,Iguchi2023,Hu2025,Lee2011,Shen2020,Malaeb2012,Grinenko2017,Terashima2010,Terashima2013,Hardy2013,Hardy2016}, where the electron pocket near the $M$ point is replaced by four propeller-shaped hole pockets at about $x=0.7$ (Fig.~\ref{figure1}(c)-(h)). Eventually, the spin fluctuations around the perfectly nesting wavevevtor ${\boldsymbol{Q}}$ in the optimal doping split into two incommensurate peaks at ${\boldsymbol{Q}}_{1,2}$ due to the mismatch of pocket sizes (Fig.~\ref{figure1}(f) and (g))\cite{Ding2008, Okazaki2012,Lee2011,Castellan2011,Horigane2016,Lee2016}. Such incommensurate spin fluctuations show a strong hole doping dependence especially in the superconducting state \cite{Castellan2011,Horigane2016,Lee2016}. Particularly in KFe$_2$As$_2$, where only anisotropic hole pockets are present both around $\Gamma$ and $M$ points, a strongly incommensurate spin resonance mode is observed at ${\boldsymbol{Q}}_{1,2}$, suggesting the persistence of $s_{\pm}$ pairing symmetry (Fig.~\ref{figure1}(h)). Therefore, it is crucial to elucidate these fascinating behaviors by exploring the interplay between the spin fluctuations and electronic structures in overdoped regime.

In this Letter, we report the low-energy spectrum of spin excitations in the hole overdoped Ba$_{1-x}$K$_x$Fe$_2$As$_2$ single crystals with $x=0.6$ and $T_c$ = 26 K, where the doping level is near the Lifshitz transition ($x=0.7$) (Fig.~\ref{figure1}(a) and (b))). For $x=0.6$ doping, the electron pocket at the $M$ point nearly vanishes, while the propeller-shaped hole-like Fermi surfaces at the $M$ point approach the Fermi energy ($E_F$), significantly altering the electronic density of states (Fig.~\ref{figure1}(c)-(k)). We observe a neutron spin resonance mode at energies of 14 meV (odd $L$) and 15 meV (even $L$), exhibiting quasi-two-dimensional characteristics, with broad incommensurate peaks centered around $Q=$ (1, 0) in the magnetic unit cell (or $Q= (0.5, 0.5)$ from (0, 0) to $(\pi, \pi)$ in the nuclear unit cell, see Fig.~\ref{figure1}(f) and (i))). A comparison of the spin excitation spectra between the superconducting and normal states reveals that the excitations evolve from a nearly non-dispersive profile in the normal state to one with downward dispersion in the superconducting state, forming a clear downward dispersion of the resonance. This behavior contrasts sharply with that observed in the optimally doped Ba$_{1-x}$K$_x$Fe$_2$As$_2$\cite{ZhangCL2011,Zhang2018,Xie2021}, suggesting that in the overdoped regime, the spin fluctuations are strongly influenced by the itinerant electrons associated with the hole pockets near the $M$ point, even though these pockets do not actually cross the Fermi level. Such modifications in Fermi surface topology drive significant changes in the physical properties of the system, especially in the spin fluctuations and electronic behaviors. Therefore, we propose a unified picture to describe the resonance dispersions which are determined by the sign-change of Fermi velocities between two nesting bands closed to the Fermi level in the superconducting state below $T_c$.

High-quality single crystals of overdoped Ba$_{0.4}$K$_{0.6}$Fe$_2$As$_2$ ($T_c = 26 \pm 2$ K) were synthesized using the self-flux method \cite{Luo2008,Luo2009}. For neutron scattering experiments, approximately 4 grams of crystals were coaligned on aluminum plates with hydrogen-free glue. The experiments were performed on the thermal triple-axis spectrometer EIGER at the Swiss Spallation Neutron Source, Paul Scherrer Institut, Switzerland, and the Taipan spectrometer at the Australian Centre for Neutron Scattering, ANSTO, Australia. The scattering plane [$H$, 0, $L$] was defined in reciprocal lattice units (r.l.u.) as ${\boldsymbol{Q}} = (H, K, L) = (q_x a/2\pi, q_y b/2\pi, q_z c/2\pi)$, where the pseudo-orthorhombic magnetic unit cell parameters are $a$ = $b$ = 5.2 \text{\AA}\ and $c$ = 13.22 \text{\AA}. The high-resolution angle-resolved photoemission (ARPES) measurements were carried out on our laboratory system equipped with a Scienta DA30 electron energy analyzer. We use a helium I resonance line as the light source which provides a photon energy of $h\nu$ =21.218 eV. The energy resolution was set at 10 meV and the angular resolution was 0.3$^{\circ}$. As shown in Fig.~\ref{figure1}(i)-(k), the ARPES results clearly demonstrate the fermiology and band structure similar to the calculated results using density functional theory (DFT) with the plane wave basis set approach implemented in Quantum ESPRESSO package (Fig.~\ref{figure1}(d) and (g)) \cite{Giannozzi2017,Perdew1996}. The superconducting gap obtained on the hole pockets around $\Gamma$ point follows a linear relation with $T_c$ ($\Delta/k_BT_c=2.15$) along with other dopings in Ba$_{1-x}$K$_x$Fe$_2$As$_2$ \cite{Wu2024,Xu2013,Nakayama2011,Cai2021}.

\begin{figure}[htbp] \centering
\includegraphics[width=0.48\textwidth]{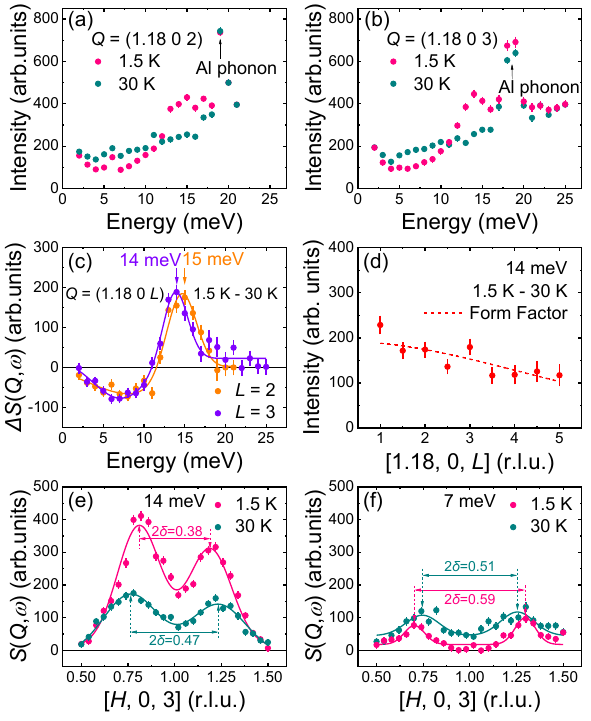}
\caption{Neutron spin resonance in Ba$_{0.4}$K$_{0.6}$Fe$_2$As$_2$. (a), (b) Energy scans at two incommensurate wave vectors: ${\boldsymbol{Q}}$ =  (1.18, 0, 2) and (1.18, 0, 3), measured below and above $T_c$. Anomalous data points are originate from aluminum phonon contributions. (c) Temperature difference plot based on the data in (a) and (b), with arrows marking the resonance energies at ${\boldsymbol{Q}}$= (1.18, 0, 2) (15 meV) and  ${\boldsymbol{Q}}$ = (1.18, 0, 3) (14 meV). (d) $L$-modulation of the resonance, where the dashed line represents the square of the magnetic form factor of Fe$^{2+}$ after normalizing to the intensity. (e), (f) Constant-energy scans at $E=$ 14 meV and 7 meV along the ($H$, 0, 3) direction, taken below and above $T_c$. Solid lines represent two-peak Gaussian fits to the data with the incommensurability $\delta$.}
\label{figure2}
\end{figure}

\begin{figure}[htbp] \centering
 \includegraphics[width=0.5\textwidth]{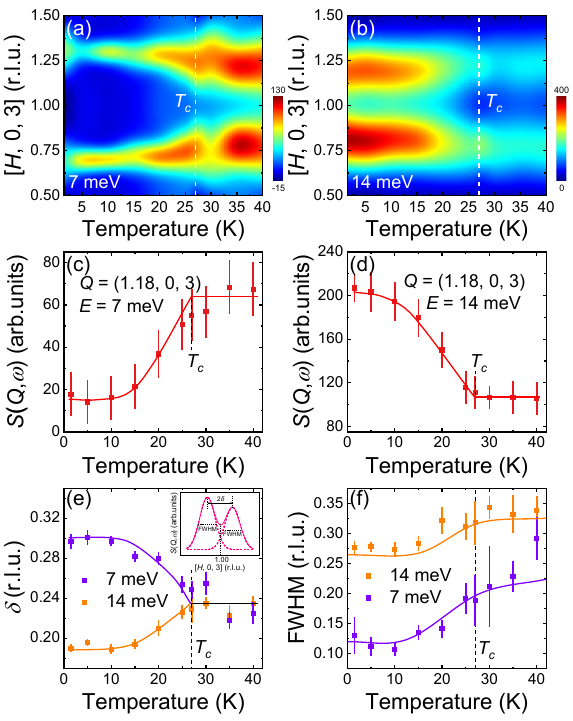}
\caption{Temperature dependence of spin excitations in Ba$_{0.4}$K$_{0.6}$Fe$_2$As$_2$. (a), (b) Spin excitations along the ($H$, 0, 3) direction at $E=$ 7 meV and 14 meV, respectively. (c), (d) Temperature dependence of spin excitations at fixed ${\boldsymbol{Q}}$ = (1.18, 0, 3) with $E=$ 7 meV and 14 meV. (e, f) Extracted incommensurability ($\delta$) and full width at half maximum (FWHM) as a function of temperature for energy transfers of 7 meV and 14 meV. Solid lines in (c), (d), (e) and (f) are guides to the eyes. The definition of $\delta$ and FWHM are illustrated in the inset of (e).
}
\label{figure3}
\end{figure}

Fig. \ref{figure2} presents the results of neutron spin resonance mode in Ba$_{0.4}$K$_{0.6}$Fe$_2$As$_2$. Fig.~\ref{figure2}(a) and (b) show the raw data collected below and above $T_c$ for odd and even $L$ values, respectively. The choice of $L$ values is critical for extracting mode energy and intensity, which can be estimated from the magnetic interaction associated with the distance between the FeAs layers\cite{Lee2013,Wa_er_2019,Hong2023}. In Ba$_{0.4}$K$_{0.6}$Fe$_2$As$_2$, the difference in mode energy and intensity between odd and even $L$ values is minimal, as illustrated in Fig.~\ref{figure2}(c). The mode energy is 14 meV for odd $L$ values and 15 meV for even $L$ values at the incommensurate wave vector (1.18, 0, $L$), with maximum intensity depletion occurring around 7 meV. Given the small discrepancy in the resonance between $L$ = 2 and $L$ = 3, we selected $L$ = 3 for further investigation. Notably, the mode energy remains identical to that of optimally doped Ba$_{1-x}$K$_{x}$Fe$_2$As$_2$\cite{ZhangCL2011,Zhang2018,Xie2021}, which peaks at a commensurate $H$ position, despite a 30$\%$ reduction in $T_c$. The $L$-modulation of the resonance follows the square of the magnetic form factor of Fe$^{2+}$, as shown in Fig.~\ref{figure2}(d), confirming its magnetic origin. The selection of incommensurate positions, essential for determining the resonant energy, is based on a combination of energy and ${\boldsymbol{Q}}$ scans, where the resonant signal is maximized, as shown in Fig.~\ref{figure2}(e). Fig.~\ref{figure2}{(e) and (f) present typical energy and momentum scans below and above $T_c$. For the resonant energy $E$ = 14 meV, the incommensurability decreases in the superconducting state, while for $E$ = 7 meV in the depletion region, it increases. This behavior can be attributed to the opening of superconducting gaps, which affects the distribution of pairing electrons near $E_F$ (Fig.~\ref{figure1}(j) and (k)).

\begin{figure}[htbp] \centering
\includegraphics[width=0.48\textwidth]{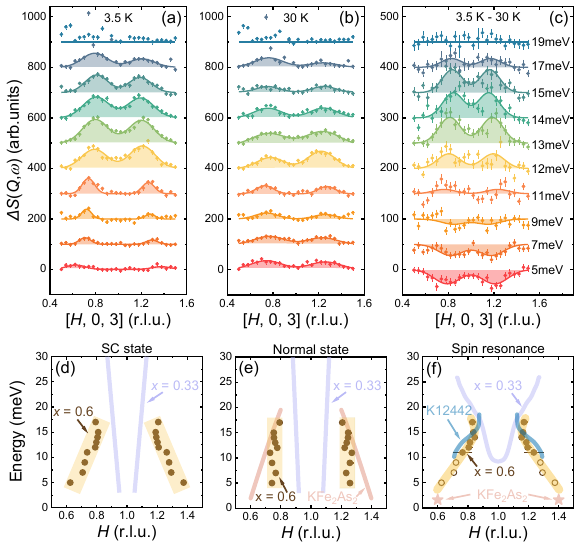}
\caption{Low-energy spin excitations in Ba$_{1-x}$K$_x$Fe$_2$As$_2$. (a), (b) Low-energy spin excitations in Ba$_{0.4}$K$_{0.6}$Fe$_2$As$_2$ along the ($H$, 0, 3) direction in both the normal and superconducting states from 5 meV to 19 meV. Solid lines represent two-peak Gaussian fits to the data, and the shaded area highlights the spin excitation signal. (c) Temperature difference plots of low-energy excitations. (d), (e) Comparison of low-energy spin excitations in Ba$_{1-x}$K$_x$Fe$_2$As$_2$ ($x$ = 0.33, 1 and from this work, 0.6, extracted from the fits in (a) and (b)) in the normal and superconducting states along the in-plane $H$ direction\cite{Xie2021,Lee2011,Shen2020,Wang2013}.  (f) Comparison of resonance dispersion in Ba$_{1-x}$K$_x$Fe$_2$As$_2$ ($x$ = 0.33, 0.6 and 1) and KCa$_2$Fe$_4$As$_4$F$_2$ (K12442), with open circles for $x$ = 0.6 representing negative net intensity.
}
\label{figure4}
\end{figure}

To explore the effects of superconductivity on low-energy spin excitations, we have performed temperature-dependent measurements at the wave vector (1.18, 0, 3) with energy transfers of 7 meV and 14 meV, as illustrated in Fig. \ref{figure3}. The results, displayed in Figs.~\ref{figure3}(a) and (b), reveal a significant suppression of intensity at 7 meV in the superconducting state. This suppression can be attributed to the redistribution of spectral weight caused by the opening of the superconducting gap, which shifts the density of states to higher energies, as evidenced by the enhanced intensity at 14 meV. The dynamic structure factor, $S({\boldsymbol{Q}}, \omega)$, exhibits an order parameter-like temperature dependence (Figs.~\ref{figure3}(c) and (d)), reinforcing the connection between the observed spin excitations and superconductivity. The different temperature evolution below $T_c$  of the incommensurability shown in ~\ref{figure3}(e)  possibly originates from the redistribution of electronic states in the momentum space, likely reflecting changes in interband scattering processes. The narrowing of peak widths at both energies in the superconducting state (Fig. ~\ref{figure3}(f)) suggests enhanced localization of magnetic scattering, consistent with the condensation of Cooper pairs.

To investigate the impact of electronic structure changes near the $M$ point on the dispersion of spin excitations in Ba$_{0.4}$K$_{0.6}$Fe$_2$As$_2$, we have also performed ${\boldsymbol{Q}}$-scans from $E=$ 5 to 19 meV. The corresponding results, shown in Fig. \ref{figure4}, are compared with those from optimally doped Ba$_{0.67}$K$_{0.33}$Fe$_2$As$_2$ and fully overdoped KFe$_2$As$_2$\cite{Xie2021,Zhang2018,Wang2013,Lee2011,Shen2020}. Fig.~\ref{figure4}(a) and (b) present low-energy spin fluctuations below and above $T_c$. The incommensurability ($\delta$) at various energies, determined by fitting with a two-peak Gaussian function, is summarized in Figs. ~\ref{figure4}(d) and (e). The net intensity and dispersion of resonance are shown in Figs. ~\ref{figure4}(c) and (f).

In Ba$_{0.4}$K$_{0.6}$Fe$_2$As$_2$, the normal-state spin excitations at the incommensurate wave vectors exhibit minimal dispersion. Upon entering the superconducting state, these excitations develop a downward-dispersive behavior. This contrasts sharply with the optimally doped Ba$_{0.67}$K$_{0.33}$Fe$_2$As$_2$, where the spin fluctuations are centered at the commensurate vector ${\boldsymbol{Q}}$ = (1, 0) and exhibit upward dispersion both below and above $T_c$ (Figs.~\ref{figure4}(d) and (e))\cite{Xie2021,Zhang2018,Wang2013}. The modification in low-energy fluctuations in Ba$_{0.4}$K$_{0.6}$Fe$_2$As$_2$ may be driven by the Lifshitz transition at the $M$ point. This interpretation is further supported by the large incommensurability observed in KFe$_2$As$_2$ due to the strongly imperfect nesting between hole pockets near the $\Gamma$ and $M$ points(Fig.~\ref{figure1}(h))\cite{Lee2016,Shen2020,Wu2024}. Notably, the resonance energy ($E_r$) in the $x=0.6$ compound keeps the same with $x=0.33$, resulting in $E_r / k_B T_c \approx 6.7$.  This deviates from the typical $E_r / k_B T_c \approx 4.9$ observed in other FeSCs but close to the  $E_r / k_B T_c \approx 5.8$ in cuprates\cite{Xie2018,Eschrig2006}, which could be a consequence from the strong-coupling Cooper pairs in hole-overdoped compounds similar to the case in KCa$_2$Fe$_4$As$_4$F$_2$ ($\Delta_{tot} / k_B T_c \approx 6$)\cite{Kim2013,Xie2021, Zhang2018,Hong2020,Fujita2012}. Here the estimated $\Delta_{tot} / k_B T_c\approx 9$ for Ba$_{0.4}$K$_{0.6}$Fe$_2$As$_2$ ($\Delta_{tot} \approx 10.3$ meV), by supposing the proportional relation between gap and $T_c$(inset of Fig.~\ref{figure1}(a)) \cite{Xie2021}. Moreover, both the $E_r$ over than $2 \Delta$ and the downward dispersion of resonance recall the case in KCa$_2$Fe$_4$As$_4$F$_2$ (K12442)(Fig.~\ref{figure4}(f)), which defy explanation by the conventional spin-exciton scenario under the $s^\pm$-pairing\cite{Maier2009,Das2011,Kim2013}.

It is intriguing to correlate the Fermi surface topologies with low-energy spin excitations in the overdoped Ba$_{1-x}$K$_x$Fe$_2$As$_2$. The downward dispersion of resonance may be due to the approach of hole pockets near $E_F$ at $M$ point, since the gap opening will expose the hole bands below $E_F$ but push away the electron bands above $E_F$ (dashed lines in Figs.~\ref{figure1}(c) and (e)). As the hole concentration increases further to $x = 1$ (namely KFe$_2$As$_2$), the Fermi surfaces near the $M$ point are dominated by four propeller-like hole pockets\cite{Xu2013,Wu2024}. Although the superconducting gaps become nodal-like in these hole pockets, the pairing symmetry is still of the $s^\pm$-type\cite{Lee2016,Shen2020,Wu2024}. In this case, the downward dispersion of resonance could be a consequence from size-mismatched hole to hole scattering, where the Fermi velocity ($V_F$) does not change sign between two nesting bands. In those optimally hole or electron doped compounds, the resonance mode is contributed by the scattering between hole and electron bands, thus its dispersion should be upward due to the sign change of $V_F$ \cite{Kim2013,Xie2021,Zhang2018}. Interestingly, such simple picture can be also applied to those iron chalcogenides with only electron pockets, where the spin excitations show an hour-glass type of dispersion in the superconducting state similar to the hole-type cuprates \cite{Wo2025,Sidis2007}. Therefore, we can establish an universal picture of the resonance dispersion in FeSCs, it is solely determined by the sign-change of Fermi velocities between two nesting bands after gap opening below $T_c$. This means the resonance dispersion does not have to follow the gap distribution in the momentum space, but it is intimately connected to the band structure near $E_F$. In most cases $E_r$ could be related to  $\Delta$ and $T_c$, but not always follow the linear scaling law when the interband scattering changes below $T_c$\cite{Wo2025,Sidis2007,Scalapino2012}.

To conclude, we used inelastic neutron scattering to investigate the low-energy spin excitations in the overdoped Ba$_{0.4}$K$_{0.6}$Fe$_2$As$_2$ near a Lifshitz transition. We identified a spin resonance mode with a peak energy $E_r$ similar to that found in optimally doped Ba$_{0.67}$K$_{0.33}$Fe$_2$As$_2$, despite a significantly reduced $T_c$. In contrast to the commensurate resonance mode and upward in-plane dispersion observed at the optimal doping, the resonance in the overdoped sample emerges at incommensurate wave vectors and exhibits a pronounced downward dispersion. These results challenge the prevailing view of the resonance as a magnetic exciton confined by the superconducting gaps, suggesting that its properties are strongly influenced by the changes in the Fermi surface topology. By establishing a unified picture of the resonance dispersions determined by the sign-change of Fermi velocities, our results underscore the pivotal role of Fermi surface evolution in driving the spin dynamics and electronic structure of FeSCs. Further experimental and theoretical work is essential to disentangle the complex interplay between multiband scattering and unconventional superconductivity, offering new insights into the roles of these low-energy collective excitations in the superconducting pairing mechanism.

\begin{acknowledgments}
This work is supported by the National Key Research and Development Program of China (Grant Nos. 2023YFA1406100, 2018YFA0704200, 2022YFA1403400 and 2021YFA1400400), the National Natural Science Foundation of China (Grant Nos. 11822411 and 12274444), the Strategic Priority Research Program (B) of the CAS (Grant Nos. XDB25000000 and XDB33000000) and K. C. Wong Education Foundation (GJTD-2020-01). AP and HG acknowledge the Computer Division at RRCAT for providing the scientific computing facilities. Financial support for this study was provided to AP by HBNI-RRCAT and MPCST under the FTYS program. This work is based on neutron scattering experiments performed at the Swiss Spallation Neutron Source (SINQ), Paul Scherrer Institut, Villigen, Switzerland (Proposal No. 20212794), and the Australian Centre for Neutron Scattering (ACNS), Australian Nuclear Science and Technology Organisation, Australia (Proposal Nos. P9850 and P9882).
\end{acknowledgments}

\bibliography{BKFA_overdoped}

\providecommand{\noopsort}[1]{}\providecommand{\singleletter}[1]{#1}%
\providecommand{\newblock}{}
\begin{thebibliography}{10}
\expandafter\ifx\csname url\endcsname\relax
  \def\url#1{{\tt #1}}\fi
\expandafter\ifx\csname urlprefix\endcsname\relax\def\urlprefix{URL }\fi
\providecommand{\eprint}[2][]{\url{#2}}

\bibitem{Scalapino2012}
Scalapino D~J 2012 {\em Rev. Mod. Phys.\/} {\bf 84}(4) 1383--1417
  \urlprefix\url{https://link.aps.org/doi/10.1103/RevModPhys.84.1383}

\bibitem{Fernandes2022}
Fernandes R~M, Coldea A~I, Ding H, Fisher I~R, Hirschfeld P~J and Kotliar G
  2022 {\em Nature\/} {\bf 601} 35--44
  \urlprefix\url{https://doi.org/10.1038/s41586-021-04073-2}

\bibitem{Mazin2008}
Mazin I~I, Singh D~J, Johannes M~D and Du M~H 2008 {\em Phys. Rev. Lett.\/}
  {\bf 101}(5) 057003
  \urlprefix\url{https://link.aps.org/doi/10.1103/PhysRevLett.101.057003}

\bibitem{Kuroki2008}
Kuroki K, Onari S, Arita R, Usui H, Tanaka Y, Kontani H and Aoki H 2008 {\em
  Phys. Rev. Lett.\/} {\bf 101}(8) 087004
  \urlprefix\url{https://link.aps.org/doi/10.1103/PhysRevLett.101.087004}

\bibitem{Si2008}
Si Q and Abrahams E 2008 {\em Phys. Rev. Lett.\/} {\bf 101}(7) 076401
  \urlprefix\url{https://link.aps.org/doi/10.1103/PhysRevLett.101.076401}

\bibitem{Seo2008}
Seo K, Bernevig B~A and Hu J 2008 {\em Phys. Rev. Lett.\/} {\bf 101}(20) 206404
  \urlprefix\url{https://link.aps.org/doi/10.1103/PhysRevLett.101.206404}

\bibitem{Dai2015}
Dai P 2015 {\em Rev. Mod. Phys.\/} {\bf 87}(3) 855--896
  \urlprefix\url{https://link.aps.org/doi/10.1103/RevModPhys.87.855}

\bibitem{Zhang2011}
Zhang Y, Yang L~X, Xu M, Ye Z~R, Chen F, He C, Xu H~C, Jiang J, Xie B~P, Ying
  J~J {\em et~al.\/} 2011 {\em Nat. Mater.\/} {\bf 10}(4) 273--277
  \urlprefix\url{https://doi.org/10.1038/nmat2981}

\bibitem{Qian2011}
Qian T, Wang X, Jin W, Zhang P, Richard P, Xu G, Dai X, Fang Z, Guo J, Chen X
  and Ding H 2011 {\em Phys. Rev. Lett.\/} {\bf 106}(18) 187001
  \urlprefix\url{https://link.aps.org/doi/10.1103/PhysRevLett.106.187001}

\bibitem{Liu2012}
Liu D, Zhang W, Mou D, He J, Ou Y, Wang Q, Li Z, Wang L, Zhao L, He S, Peng Y,
  Liu X, Chen C, Yu L, Liu G, Dong X, Zhang J, Chen C, Xu Z, Hu J, Chen X, Ma
  X, Xue Q and Zhou X~J 2012 {\em Nat. Commun.\/} {\bf 3} 931
  \urlprefix\url{https://doi.org/10.1038/ncomms1946}

\bibitem{Zhao2016}
Zhao L, Liang A, Yuan D, Hu Y, Liu D, Huang J, He S, Shen B, Xu Y, Liu X, Yu L,
  Liu G, Zhou H, Huang Y, Dong X, Zhou F, Liu K, Lu Z, Zhao Z, Chen C, Xu Z and
  Zhou X~J 2016 {\em Nat. Commun.\/} {\bf 7} 10608
  \urlprefix\url{https://doi.org/10.1038/ncomms10608}

\bibitem{Wo2025}
Wo H, Pan B, Hu D, Feng Y, Christianson A~D and Zhao J 2025 {\em Phys. Rev.
  Lett.\/} {\bf 134}(1) 016501
  \urlprefix\url{https://link.aps.org/doi/10.1103/PhysRevLett.134.016501}

\bibitem{Maier2011}
Maier T~A, Graser S, Hirschfeld P~J and Scalapino D~J 2011 {\em Phys. Rev. B\/}
  {\bf 83}(10) 100515
  \urlprefix\url{https://link.aps.org/doi/10.1103/PhysRevB.83.100515}

\bibitem{Wang2011}
Wang F, Yang F, Gao M, Lu Z, Xiang T and Lee D 2011 {\em Europhys. Lett.\/}
  {\bf 93} 57003 \urlprefix\url{https://doi.org/10.1209/0295-5075/93/57003}

\bibitem{Dai2012}
Dai P, Hu J and Dagotto E 2012 {\em Nat. Phys.\/} {\bf 8}(10) 709--718
  \urlprefix\url{https://doi.org/10.1038/nphys2438}

\bibitem{Wang2013}
Wang M, Zhang C, Lu X, Tan G, Luo H, Song Y, Wang M, Zhang X, Goremychkin E~A,
  Perring T~G, Maier T~A, Yin Z, Haule K, Kotliar G and Dai P 2013 {\em Nat.
  Commun.\/} {\bf 4} 2874 \urlprefix\url{https://doi.org/10.1038/ncomms3874}

\bibitem{Pan2017}
Pan B, Shen Y, Hu D and Zhao J 2017 {\em Nat. Commun.\/} {\bf 8} 123
  \urlprefix\url{https://doi.org/10.1038/s41467-017-00162-x}

\bibitem{Friemel2012}
Friemel G, Park J~T, Maier T~A, Tsurkan V, Li Y, Deisenhofer J, Krug~von Nidda
  H~A, Loidl A, Ivanov A, Keimer B and Inosov D~S 2012 {\em Phys. Rev. B\/}
  {\bf 85}(14) 140511
  \urlprefix\url{https://link.aps.org/doi/10.1103/PhysRevB.85.140511}

\bibitem{Wang2012}
Wang M, Li C, Abernathy D~L, Song Y, Carr S~V, Lu X, Li S, Yamani Z, Hu J,
  Xiang T and Dai P 2012 {\em Phys. Rev. B\/} {\bf 86}(2) 024502
  \urlprefix\url{https://link.aps.org/doi/10.1103/PhysRevB.86.024502}

\bibitem{Bohmer2015}
B{\"o}hmer A~E, Hardy F, Wang L, Wolf T, Schweiss P and Meingast C 2015 {\em
  Nat. Commun.\/} {\bf 6} 7911
  \urlprefix\url{https://doi.org/10.1038/ncomms8911}

\bibitem{Wu2024}
Wu D, Jia J, Yang J, Hong W, Shu Y, Miao T, Yan H, Rong H, Ai P, Zhang X, Yin
  C, Liu J, Chen H, Yang Y, Peng C, Li C, Zhang S, Zhang F, Yang F, Wang Z,
  Zong N, Liu L, Li R, Wang X, Peng Q, Mao H, Liu G, Li S, Chen Y, Luo H, Wu X,
  Xu Z, Zhao L and Zhou X~J 2024 {\em Nat. Phys.\/} {\bf 20} 571--578
  \urlprefix\url{https://doi.org/10.1038/s41567-023-02348-1}

\bibitem{Xu2013}
Xu N, Richard P, Shi X, van Roekeghem A, Qian T, Razzoli E, Rienks E, Chen G,
  Ieki E, Nakayama K, Sato T, Takahashi T, Shi M and Ding H 2013 {\em Phys.
  Rev. B\/} {\bf 88}(22) 220508
  \urlprefix\url{https://link.aps.org/doi/10.1103/PhysRevB.88.220508}

\bibitem{Nakayama2011}
Nakayama K, Sato T, Richard P, Xu Y, Kawahara T, Umezawa K, Qian T, Neupane M,
  Chen G~F, Ding H and Takahashi T 2011 {\em Phys. Rev. B\/} {\bf 83}(2) 020501
  \urlprefix\url{https://link.aps.org/doi/10.1103/PhysRevB.83.020501}

\bibitem{Cai2021}
Cai Y, Huang J, Miao T, Wu D, Gao Q, Li C, Xu Y, Jia J, Wang Q, Huang Y, Liu G,
  Zhang F, Zhang S, Yang F, Wang Z, Peng Q, Xu Z, Zhao L and Zhou X~J 2021 {\em
  Sci. Bull.\/} {\bf 66} 1839--1848
  \urlprefix\url{https://doi.org/10.1016/j.scib.2021.05.015}

\bibitem{Kihou2016}
Kihou K, Saito T, Fujita K, Ishida S, Nakajima M, Horigane K, Fukazawa H,
  Kohori Y, Uchida S~I, Akimitsu J, Iyo A and Eisaki H 2016 {\em J. Phys. Soc.
  Jpn.\/} {\bf 85}(3) 034718
  \urlprefix\url{https://doi.org/10.7566/JPSJ.85.034718}

\bibitem{Thomale2011}
Thomale R, Platt C, Hanke W, Hu J and Bernevig B~A 2011 {\em Phys. Rev.
  Lett.\/} {\bf 107}(11) 117001
  \urlprefix\url{https://link.aps.org/doi/10.1103/PhysRevLett.107.117001}

\bibitem{Grinenko2020}
Grinenko V, Sarkar R, Kihou K, Lee C, Morozov I, Aswartham S, Büchner B,
  Chekhonin P, Skrotzki W, Nenkov K, Hühne R, Nielsch K, Drechsler S, Vadimov
  V~L, Silaev M~A, Volkov P~A, Eremin I, Luetkens H and Klauss H 2020 {\em Nat.
  Phys.\/} {\bf 16} 789--794
  \urlprefix\url{https://doi.org/10.1038/s41567-020-0886-9}

\bibitem{Bartl2025}
Bärtl F, Stegani N, Caglieris F, Shipulin I, Li Y, Hu Q, Zheng Y, Yim C,
  Luther S, Wosnitza J {\em et~al.\/} 2025 {\em arXiv preprint
  arXiv:2501.11936\/} \urlprefix\url{https://doi.org/10.48550/arXiv.2501.11936}

\bibitem{Iguchi2023}
Iguchi Y, Shi R~A, Kihou K, Lee C~H, Barkman M, Benfenati A~L, Grinenko V,
  Babaev E and Moler K~A 2023 {\em Science\/} {\bf 380} 1244--1247
  (\textit{Preprint}
  \eprint{https://www.science.org/doi/pdf/10.1126/science.abp9979})
  \urlprefix\url{https://www.science.org/doi/abs/10.1126/science.abp9979}

\bibitem{Hu2025}
Hu Q, Zheng Y, Xu H, Deng J, Liang C, Yang F, Wang Z, Grinenko V, Lv B, Ding H
  and Yim C~M 2025 {\em Nat. Commun.\/} {\bf 16} 253
  \urlprefix\url{https://doi.org/10.1038/s41467-024-55368-7}

\bibitem{Lee2011}
Lee C~H, Kihou K, Kawano-Furukawa H, Saito T, Iyo A, Eisaki H, Fukazawa H,
  Kohori Y, Suzuki K, Usui H, Kuroki K and Yamada K 2011 {\em Phys. Rev.
  Lett.\/} {\bf 106}(6) 067003
  \urlprefix\url{https://link.aps.org/doi/10.1103/PhysRevLett.106.067003}

\bibitem{Shen2020}
Shen S, Zhang X, Wo H, Shen Y, Feng Y, Schneidewind A, Čermák P, Wang W and
  Zhao J 2020 {\em Phys. Rev. Lett.\/} {\bf 124}(1) 017001
  \urlprefix\url{https://link.aps.org/doi/10.1103/PhysRevLett.124.017001}

\bibitem{Malaeb2012}
Malaeb W, Shimojima T, Ishida Y, Okazaki K, Ota Y, Ohgushi K, Kihou K, Saito T,
  Lee C~H, Ishida S, Nakajima M, Uchida S, Fukazawa H, Kohori Y, Iyo A, Eisaki
  H, Chen C, Watanabe S, Ikeda H and Shin S 2012 {\em Phys. Rev. B\/} {\bf
  86}(16) 165117
  \urlprefix\url{https://link.aps.org/doi/10.1103/PhysRevB.86.165117}

\bibitem{Grinenko2017}
Grinenko V, Materne P, Sarkar R, Luetkens H, Kihou K, Lee C~H, Akhmadaliev S,
  Efremov D~V, Drechsler S~L and Klauss H~H 2017 {\em Phys. Rev. B\/} {\bf
  95}(21) 214511
  \urlprefix\url{https://link.aps.org/doi/10.1103/PhysRevB.95.214511}

\bibitem{Terashima2010}
Terashima T, Kimata M, Kurita N, Satsukawa H, Harada A, Hazama K, Imai M, Sato
  A, Kihou K, Lee C, Kito H, Eisaki H, Iyo A, Saito T, Fukazawa H, Kohori Y,
  Harima H and Uji S 2010 {\em J. Phys. Soc. Jpn.\/} {\bf 79}(5) 053702
  \urlprefix\url{https://doi.org/10.1143/JPSJ.79.053702}

\bibitem{Terashima2013}
Terashima T, Kurita N, Kimata M, Tomita M, Tsuchiya S, Imai M, Sato A, Kihou K,
  Lee C, Kito H, Eisaki H, Iyo A, Saito T, Fukazawa H, Kohori Y, Harima H and
  Uji S 2013 {\em Phys. Rev. B\/} {\bf 87}(22) 224512
  \urlprefix\url{https://link.aps.org/doi/10.1103/PhysRevB.87.224512}

\bibitem{Hardy2013}
Hardy F, Böhmer A~E, Aoki D, Burger P, Wolf T, Schweiss P, Heid R, Adelmann P,
  Yao Y~X, Kotliar G, Schmalian J and Meingast C 2013 {\em Phys. Rev. Lett.\/}
  {\bf 111}(2) 027002
  \urlprefix\url{https://link.aps.org/doi/10.1103/PhysRevLett.111.027002}

\bibitem{Hardy2016}
Hardy F, Böhmer A~E, de' Medici L, Capone M, Giovannetti G, Eder R, Wang L, He
  M, Wolf T, Schweiss P, Heid R, Herbig A, Adelmann P, Fisher R~A and Meingast
  C 2016 {\em Phys. Rev. B\/} {\bf 94}(20) 205113
  \urlprefix\url{https://link.aps.org/doi/10.1103/PhysRevB.94.205113}

\bibitem{Ding2008}
Ding H, Richard P, Nakayama K, Sugawara K, Arakane T, Sekiba Y, Takayama A,
  Souma S, Sato T, Takahashi T, Wang Z, Dai X, Fang Z, Chen G~F, Luo J~L and
  Wang N~L 2008 {\em Europhys. Lett.\/} {\bf 83} 47001
  \urlprefix\url{https://doi.org/10.1209/0295-5075/83/47001}

\bibitem{Okazaki2012}
Okazaki K, Ota Y, Kotani Y, Malaeb W, Ishida Y, Shimojima T, Kiss T, Watanabe
  S, Chen C~T, Kihou K, Lee C~H, Iyo A, Eisaki H, Saito T, Fukazawa H, Kohori
  Y, Hashimoto K, Shibauchi T, Matsuda Y, Yamashita M, Iwasawa H, Ikeda H and
  Shin S 2012 {\em Science\/} {\bf 337} 1314--1317
  \urlprefix\url{https://doi.org/10.1126/science.1222793}

\bibitem{Castellan2011}
Castellan J~P, Rosenkranz S, Goremychkin E~A, Chung D~Y, Todorov I~S,
  Kanatzidis M~G, Eremin I, Knolle J, Chubukov A~V, Maiti S, Norman M~R, Weber
  F, Claus H, Guidi T, Bewley R~I and Osborn R 2011 {\em Phys. Rev. Lett.\/}
  {\bf 107}(17) 177003
  \urlprefix\url{https://link.aps.org/doi/10.1103/PhysRevLett.107.177003}

\bibitem{Horigane2016}
Horigane K, Kihou K, Fujita K, Kajimoto R, Ikeuchi K, Ji S, Akimitsu J and Lee
  C~H 2016 {\em Sci. Rep.\/} {\bf 6} 33303
  \urlprefix\url{https://doi.org/10.1038/srep33303}

\bibitem{Lee2016}
Lee C~H, Kihou K, Park J~T, Horigane K, Fujita K, Waßer F, Qureshi N, Sidis Y,
  Akimitsu J and Braden M 2016 {\em Sci. Rep.\/} {\bf 6} 23424
  \urlprefix\url{https://doi.org/10.1038/srep23424}

\bibitem{ZhangCL2011}
Zhang C, Wang M, Luo H, Wang M, Liu M, Zhao J, Abernathy D~L, Maier T~A, Marty
  K, Lumsden M~D, Chi S, Chang S, RodriguezRivera J~A, Lynn J~W, Xiang T, Hu J
  and Dai P 2011 {\em Sci. Rep.\/} {\bf 1} 115
  \urlprefix\url{https://doi.org/10.1038/srep00115}

\bibitem{Zhang2018}
Zhang R, Wang W, Maier T~A, Wang M, Stone M~B, Chi S, Winn B and Dai P 2018
  {\em Phys. Rev. B\/} {\bf 98}(6) 060502
  \urlprefix\url{https://link.aps.org/doi/10.1103/PhysRevB.98.060502}

\bibitem{Xie2021}
Xie T, Liu C, Fennell T, Stuhr U, Li S and Luo H 2021 {\em Chin. Phys. B\/}
  {\bf 30}(12) 127402
  \urlprefix\url{https://dx.doi.org/10.1088/1674-1056/ac3651}

\bibitem{Luo2008}
Luo H, Wang Z, Yang H, Cheng P, Zhu X and Wen H 2008 {\em Supercond. Sci.
  Technol.\/} {\bf 21} 125014
  \urlprefix\url{https://dx.doi.org/10.1088/0953-2048/21/12/125014}

\bibitem{Luo2009}
Luo H, Cheng P, Wang Z, Yang H, Jia Y, Fang L, Ren C, Shan L and Wen H 2009
  {\em Physica C\/} {\bf 469} 477--484
  \urlprefix\url{https://www.sciencedirect.com/science/article/pii/S0921453409000835}

\bibitem{Giannozzi2017}
Giannozzi P, Andreussi O, Brumme T, Bunau O, Nardelli M~B, Calandra M, Car R,
  Cavazzoni C, Ceresoli D, Cococcioni M {\em et~al.\/} 2017 {\em J. Phys.:
  Condens. Matter\/} {\bf 29} 465901
  \urlprefix\url{https://dx.doi.org/10.1088/1361-648X/aa8f79}

\bibitem{Perdew1996}
Perdew J~P, Burke K and Ernzerhof M 1996 {\em Phys. Rev. Lett.\/} {\bf 77}(18)
  3865--3868
  \urlprefix\url{https://link.aps.org/doi/10.1103/PhysRevLett.77.3865}

\bibitem{Lee2013}
Lee C~H, Steffens P, Qureshi N, Nakajima M, Kihou K, Iyo A, Eisaki H and Braden
  M 2013 {\em Phys. Rev. Lett.\/} {\bf 111}(16) 167002
  \urlprefix\url{https://link.aps.org/doi/10.1103/PhysRevLett.111.167002}

\bibitem{Wa_er_2019}
Waßer F, Park J~T, Aswartham S, Wurmehl S, Sidis Y, Steffens P, Schmalzl K,
  Büchner B and Braden M 2019 {\em npj Quantum Mater.\/} {\bf 4} 59
  \urlprefix\url{http://dx.doi.org/10.1038/s41535-019-0198-4}

\bibitem{Hong2023}
Hong W, Zhou H, Li Z, Li Y, Stuhr U, Pokhriyal A, Ghosh H, Tao Z, Lu X, Hu J,
  Li S and Luo H 2023 {\em Phys. Rev. B\/} {\bf 107}(22) 224514
  \urlprefix\url{https://link.aps.org/doi/10.1103/PhysRevB.107.224514}

\bibitem{Xie2018}
Xie T, Gong D, Ghosh H, Ghosh A, Soda M, Masuda T, Itoh S, Bourdarot F,
  Regnault L, Danilkin S, Li S and Luo H 2018 {\em Phys. Rev. Lett.\/} {\bf
  120}(13) 137001
  \urlprefix\url{https://link.aps.org/doi/10.1103/PhysRevLett.120.137001}

\bibitem{Eschrig2006}
Eschrig M 2006 {\em Adv. Phys.\/} {\bf 55} 47--183
  \urlprefix\url{https://doi.org/10.1080/00018730600645636}

\bibitem{Kim2013}
Kim M~G, Tucker G~S, Pratt D~K, Ran S, Thaler A, Christianson A~D, Marty K,
  Calder S, Podlesnyak A, Bud'ko S~L, Canfield P~C, Kreyssig A, Goldman A~I and
  McQueeney R~J 2013 {\em Phys. Rev. Lett.\/} {\bf 110}(17) 177002
  \urlprefix\url{https://link.aps.org/doi/10.1103/PhysRevLett.110.177002}

\bibitem{Hong2020}
Hong W, Song L, Liu B, Li Z, Zeng Z, Li Y, Wu D, Sui Q, Xie T, Danilkin S,
  Ghosh H, Ghosh A, Hu J, Zhao L, Zhou X, Qiu X, Li S and Luo H 2020 {\em Phys.
  Rev. Lett.\/} {\bf 125}(11) 117002
  \urlprefix\url{https://link.aps.org/doi/10.1103/PhysRevLett.125.117002}

\bibitem{Fujita2012}
Fujita M, Hiraka H, Matsuda M, Matsuura M, Tranquada J~M, Wakimoto S, Xu G and
  Yamada K 2012 {\em J. Phys. Soc. Jpn.\/} {\bf 81}(1) 011007
  \urlprefix\url{https://doi.org/10.1143/JPSJ.81.011007}

\bibitem{Maier2009}
Maier T~A, Graser S, Scalapino D~J and Hirschfeld P 2009 {\em Phys. Rev. B\/}
  {\bf 79}(13) 134520
  \urlprefix\url{https://link.aps.org/doi/10.1103/PhysRevB.79.134520}

\bibitem{Das2011}
Das T and Balatsky A~V 2011 {\em Phys. Rev. Lett.\/} {\bf 106}(15) 157004
  \urlprefix\url{https://link.aps.org/doi/10.1103/PhysRevLett.106.157004}

\bibitem{Sidis2007}
Sidis Y, Pailhès S, Hinkov V, Fauqué B, Ulrich C, Capogna L, Ivanov A,
  Regnault L, Keimer B and Bourges P 2007 {\em C. R. Phys.\/} {\bf 8} 745--762
  \urlprefix\url{https://www.sciencedirect.com/science/article/pii/S163107050700165X}

\end{thebibliography}

\end{document}